\documentclass[twocolumn]{aastex62}

\usepackage{natbib}

\begin{document}

\title{Spectropolarimetric Insight into Plasma-Sheet Dynamics of a Solar Flare}

\author{Ryan J. French}
\affil{Mullard Space Science Laboratory, University College London, Dorking, RH5 6NT, UK}
\author{Philip G. Judge}
\affil{HAO, National Center for Atmospheric Research, P.O. Box 3000, Boulder CO 80307-3000, USA}
\author{Sarah A. Matthews}
\affil{Mullard Space Science Laboratory, University College London, Dorking, RH5 6NT, UK}\author{Lidia van Driel-Gesztelyi}
\affil{Mullard Space Science Laboratory, University College London, Dorking, RH5 6NT, UK}\affil{LESIA, Observatoire de Paris, Université PSL, CNRS, Sorbonne Université, Université Paris Diderot, 5 place Jules Janssen, 92190 Meudon, France}
\affil{Konkoly Observatory of the Hungarian Academy of Sciences, Budapest, Hungary}


\begin{abstract}

We examine spectropolarimetric data from the CoMP instrument, acquired during the evolution of the September 10th 2017 X8.2 solar flare on the western solar limb. CoMP captured linearly polarized light from two emission lines of \ion{Fe}{13} at 1074.7 and 1079.8 nm,
from 1.03 to 1.5 solar radii. We focus here on the hot plasma-sheet 
lying above the bright flare loops and beneath the ejected CME. The polarization has a striking and coherent spatial
structure, with unexpectedly small polarization aligned with the plasma-sheet. By elimination, we find that small-scale \textit{magnetic field} structure 
 is needed to cause such significant depolarization, and suggest that plasmoid formation during reconnection
(associated with the tearing mode instability) creates magnetic structure
on scales below instrument resolution of 6 Mm. We conclude
that polarization measurements with new coronagraphs, such as the upcoming DKIST, will further enhance our understanding of magnetic reconnection and development of turbulence in the solar corona. 

\end{abstract}

\section{Introduction}

Magnetic reconnection is thought to lie at the heart of energy release in solar flares. The earliest models based upon the dynamics of a planar current-sheet 
\citep{Sweet1958,Parker1957} gave rise to slow reconnection rates that depend on the magnetic Lundquist Number $S$ by $1/\sqrt{S}$, where $S = v_A L/\eta$. Here, $v_A$ is the Alfv\'en speed, $L$
the current-sheet half-length and $\eta$ the magnetic diffusivity. In the corona, $S \approx 10^{12}$. 
An alternative MHD model was proposed by \cite{Petschek1964}, in which fast steady-state 
reconnection takes place along a small
fraction of the current-sheet length, made possible by the inclusion of slow shocks. Petschek's model remains of interest as it was the first to yield reconnection rates fast enough to account for 
the rapid energy release observed in flares, varying instead as $1/\log S$. 

These current-sheet configurations were incorporated into the 
the standard ``CSHKP'' solar flare model (\citealp{Carmichael}; \citealp{Sturrock}; \citealp{Hirayama}; \citealp{Kopp}), where a rising flux rope causes the inflow of oppositely orientated magnetic field lines, creating between them a Sweet-Parker current-sheet in which reconnection occurs.
The standard model faces well-known fundamental challenges, related to the plasma micro-physics.
Petschek's mechanism assumes a certain large-scale steady configuration, 
but questions surround how such a configuration might occur \citep[e.g.][]{Kulsrud2011}. Furthermore, additional physics must be introduced to explain how the plasma is heated, and how electric fields capable of accelerating particles to above MeV energies are generated. \citep[e.g.][]{Benz2016}.

In recent years, attention has been drawn to 
the possible role of a tearing mode instability 
across current-sheets (or
\textit{plasmoid instability}), in explaining the onset of ``fast'' reconnection, i.e. at a rate independent of $S$, in various regimes. In magneto-hydrodynamics (MHD), analytical growth rates of the plasmoid instability were derived under conditions 
where current-sheet lengths greatly exceed their (MHD) thickness (\citealp{1990PhFlA...2.1487C,2007PhPl...14j0703L}). Such current-sheets were found to be intrinsically unstable to high-wavenumber perturbations, with growth rates greatly in excess of Alfv\'en crossing times. 
A chain of number $S^{3/8}$ ($\approx 10^4$ in the solar corona) plasmoids are formed along sheet length $2L$, each with a length $S^{1/8}$ larger than the current-sheet width $\delta= L/\sqrt{S}$. Numerical 2D simulations with $S=10^6$ have supported the general picture of disruption of reconnecting current-sheets through the plasmoid instability, creating a turbulent cascade with a power spectrum consistent with in-situ observations of plasma turbulence \citep{2018PhRvL.121p5101D}. However, theoretical work must still be guided by observations.

In this study, we present observational evidence for the presence of unobservably 
small magnetic structure, consistent with the plasmoid-fragmentation picture within a dynamically evolving current-sheet in the wake of a coronal mass ejection. We show that the magnitude of linear polarization is sensitive to unresolvable small-scale magnetic structures.

\section{The plasma-sheet associated with the September 10th 2017 flare}

In the corona, current-sheets are predicted to occur with a width of order 10 m \citep{Litvinenko}, far below the observable limit of even the best coronal instruments
($\approx 200$ km). However, rare sheets of hot plasma have been observed, associated with eruptive flares and appearing to be related to reconnection within a current-sheet \citep[e.g.][]{Liu}. 
These ``\textit{plasma}-sheets'' are elusive and notoriously difficult to identify, most readily seen above the solar limb.

Perhaps the brightest and longest-lived plasma-sheet observation to date is associated with an X8.2-class flare on September 10th 2017 (e.g. \citealp{Long}; \citealp{Warren}; \citealp{Kuridze}; \citealp{Li}, \citealp{Cheng}; \citealp{Longcope}; \citealp{Gary}; \citealp{Morosan}). 
The flare and coronal mass ejection
erupted from AR 12673 on the western solar limb, observed across the spectrum by multiple space-based and ground-based instruments. 
Fortuitously, the CoMP instrument 
obtained polarization data of the 
plasma-sheet, and although the plasma-sheet has been well studied, no analysis of its
polarized light has been published.

Important earlier studies of intensity images and spectra include that of 
\cite{Warren}. They used Extreme ultraviolet (EUV) Imaging Spectrometer (EIS) \citep{Culhane} and Atmospheric Imaging Assembly (AIA) \citep{Lemen} data to study the spectroscopic evolution and structure of the plasma-sheet. Using temperature sensitive EIS lines, they calculated a mean plasma-sheet temperature of 15-20 MK. It was deduced that the plasma-sheet must be heated by processes originating from magnetic reconnection, as is consistent with the CSHKP model. 

\cite{Li} and \cite{Warren} investigated non-thermal broadening
 of spectral lines 
within the plasma-sheet, finding non-thermal velocities as high as 200 km s$^{-1}$. The highest line widths (measuring 
velocities of plasma superposed along the line-of-sight, or ``LOS'')
were seen first at the base of the plasma-sheet, later they shifted to higher altitudes. The broad lines were hypothesized to indicate small-scale turbulent velocity fluctuations from plasmoid fragmentation during reconnection. In support of this idea, \cite{Cheng} analyzed the plasma-sheet plane-of-sky (POS) outflows and find a power-law spectrum of fluctuations in wavenumber space consistent with a turbulent cascade of energy toward smaller scales.

Thus, while there is indirect evidence for the presence of instabilities in the plasma-sheet of the September 10th 2017 flare, none of the evidence has provided clear insight into the nature of the plasma-sheet's magnetic field and its role in onset of turbulence. More direct observations of the magnetic field may be a crucial clue to our understanding magnetic reconnection in this and similar events.

\begin{figure*}
  \centering
  \includegraphics[width=16cm]{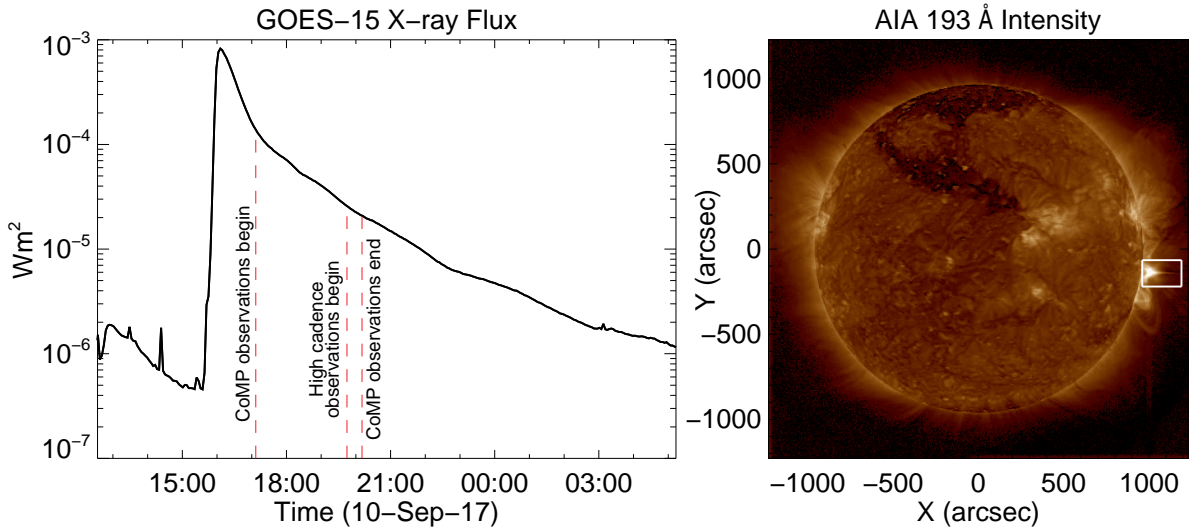}
  \caption{Left: GOES-15 X-ray flux for the flare, displaying CoMP observing times. 
  Right: Location of the plasma-sheet FOV used in this study.}
  \label{fig:event}
\end{figure*}

\section{Observations}

The September 10th 2017 flare originated from AR 12673 on the western limb, peaking at 16:06 UT. In this study we use observations from the High Altitude Observatory (HAO) Coronal Multi-channel Polarimeter (CoMP), between 17:07:50 and 20:10:36 UT. The CoMP instrument has an aperture of 20 cm and uses a coronagraph to observe the low corona from $\sim$1.03 to 1.5 \(R_\odot\). CoMP measures the intensity and linear polarization (Stokes I,Q,U) of infrared Fe XIII 1074.7 nm and 1079.8 nm lines, with a formation temperature of $\sim$1.5 MK. 48 of the 62 available observations occurring between 19:44:36 and 20:10:36 UT 
measured the Fe XIII lines centered at three wavelengths, each 
through a filter of
roughly Gaussian shape (FWHM of 1.3 \AA),
with a 4.35\arcsec\ spatial sampling and 30 second cadence. 
CoMP observing times are shown in the left panel of Figure \ref{fig:event}.

The K-Cor instrument at the Mauna Loa Solar Observatory also observed the event, measuring white light polarization ($pB$) from 1.05 to 3 \(R_\odot\) over the same observing duration as CoMP. K-Cor has a lower resolution than CoMP (spatial sampling of 5.64\arcsec) but a higher cadence of 15 s.

EUV observations by AIA onboard the Solar Dynamics Observatory (SDO) provide context, with a higher cadence ($\sim12$ seconds) and considerably higher spatial resolution (0.6\arcsec). The plasma-sheet is most visible in the 193 {\AA} passband, measuring both Fe XXIV and Fe XII emission. Given the plasma-sheet's high temperature, most of the observed emission is likely from the 20 MK Fe XXIV line.
Despite its high temperature, the plasma-sheet is also seen in cooler AIA passbands, such as 211 {\AA} (\cite{Warren}), dominated by plasma closer to 2 MK.
In AIA 193 {\AA}, the plasma-sheet is clearly visible from 16:06 to beyond 20:30 UT. Therefore, although the higher cadence CoMP observations start 2 h 38 m after the flare peak, the plasma-sheet is still visible in EUV observations during this time. This is much longer than the Alfv\'en crossing time, which is just a few minutes
for a magnetic field strength of 10G. 

\begin{figure*}
  \centering
  \includegraphics[width=14.5cm]{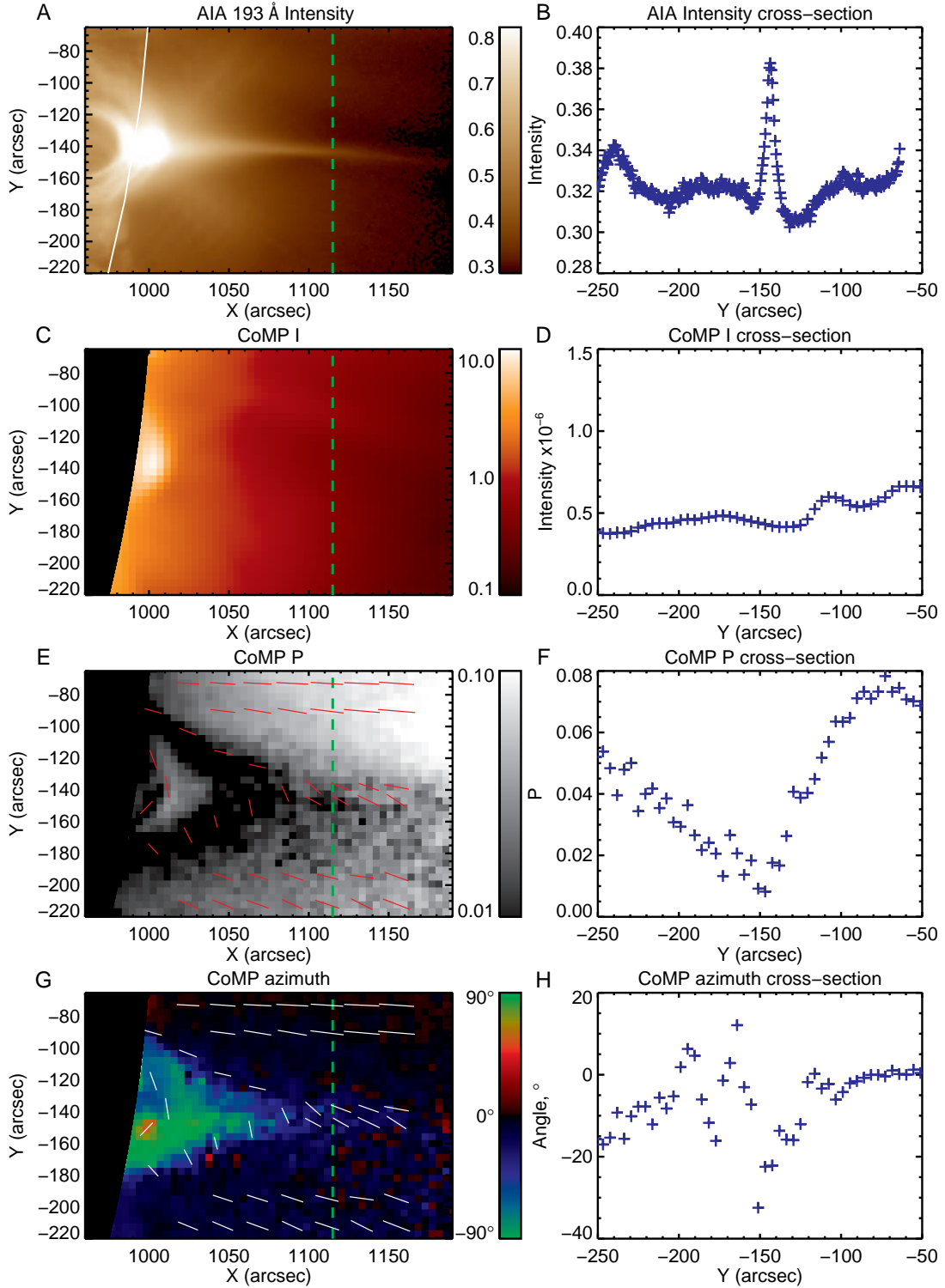}
  \caption{
  A) Normalized AIA 193 {\AA} intensity. Curved white line marks position of the CoMP occulting disk. 
  B) AIA 193{\AA} cross-section of intensity along the dashed green line in adjacent panel.
  C) CoMP 1074.7 nm $I$
  D) CoMP 1074.7 nm cross-section of $I$, along the dashed green line in adjacent panel.
  E) CoMP 1074.7 nm $P$. Red lines are polarization vectors, with length proportional to $-1/\log(P)$.
  F) CoMP 1074.7 nm cross-section of $P$, along the dashed green line in adjacent panel. 
  G) CoMP 1074.7 nm azimuth angle $\theta$, relative to the radial direction. White lines show the corresponding polarization vectors.
  H) CoMP 1074.7 nm cross section of $\theta$, along the dashed green lines in adjacent panel. CoMP images have an overlaid artificial-occulter to increase sharpness at image edge.
  }
  \label{fig:polarization}
\end{figure*}

Figure \ref{fig:polarization}A shows AIA 193 {\AA} observations of the plasma-sheet, averaged over the CoMP observing period and processed using the Multi-Gaussian Normalisation (MGN) technique (\cite{Morgan}). In this image, the plasma-sheet is seen as the bright horizontal structure, located above the saturated flare loop. A diffraction pattern from saturated intensities is also visible, as a faint cross emanating from the flaring region. The location of this field of view (FOV) is shown in Figure \ref{fig:event}B.

\section{Spectropolarimetry}

Linearly polarized radiation is created by the scattering of anisotropic radiation from the solar surface by coronal plasma. The anisotropic radiation generates unequal populations of magnetic sub-states, (\textit{atomic} polarization), dependent on the local thermal and magnetic conditions of the plasma \citep{Charvin1965}. The atomic polarization of \ion{Fe}{13} is, to within a few percent, proportional to the factor $3 \cos^2\theta_B-1$, where $\theta_B$ is the angle between the magnetic field vector and direction of the center of the incident radiation (equation 45 of \cite{CasiniJudge}, \cite{Judge2007}). Emission from a complex atom excited by anisotropic radiation must be computed from a solution to the statistical equilibrium equations. The term $3 \cos^2\theta_B-1$ is the leading order angular
factor in the radiative excitation of atomic sub-levels for each
transition. \cite{Judge2007} demonstrated that for multiple levels in an atomic model, the linear polarization arising from sub-level populations follow this term within a few percent for typical M1 transitions. This is because incident solar radiation
is strongest in the optical and infrared wavelengths where the M1 transitions
are found. The polarization can be destroyed by isotropic processes, including collisions by a sufficiently high density of thermal electrons and protons.

The emitted radiation is linearly polarized by a factor
proportional to
the amount of atomic polarization, therefore varying as
\begin{equation} \nonumber
  P \propto 3 \cos^2\theta_B-1.
  \label{eqn:linear}
\end{equation}

 For a radial magnetic field, the magnetic field vector is parallel to incident radiation ($\theta_B = 0$) and linear polarization is at a maximum. For a tangential magnetic field, atomic polarization becomes negative. The atomic polarization passes through
 zero as $3\cos^2\theta_B=1$, at the `Van Vleck' angle $\theta_B=\theta_{VV} = 54.74^{\circ}$. Because we have no prior knowledge of $\theta_B$ relative to $\theta_{VV}$, the change in sign of atomic polarization leads to
 a well-known 90$^{\circ}$ ambiguity in determining the POS projection of magnetic field direction.

CoMP measures the two components of linear polarization relative to
a fixed reference direction, as well as total unpolarized intensity (Stokes $U$, $Q$ and $I$ respectively). Combining these, we calculate fractional linear polarization through,
\begin{equation}
  P = \sqrt{U^2 + Q^2}/I. 
  \label{eqn:stokes}
\end{equation}
We can also use Stokes U and Q to calculate the azimuth angle of the polarization vector in the POS,
\begin{equation}
  \theta = \frac{1}{2}\arctan{(\frac{U}{Q})}.
  \label{eqn:azi}
\end{equation}
While $\theta$ is determined by $U$ and $Q$ measurements, the corresponding polarization vector has 
the $90^{\circ}$ ambiguity to magnetic field lines, either parallel, perpendicular, or undetermined depending on whether 
the actual (unknown) 
angle $\theta_B$ is 
greater than, smaller than or equal to $\theta_{VV}$.
The polarization vector is not a physical ``vector'' but a line with a magnitude and azimuth.

The corona is optically thin to infrared radiation. Therefore, every observation involves 
integration over the LOS. Variations in $\theta_B$ along the LOS lead
to a superposition of different polarization vectors (weighted by the local plasma density), causing a reduction of $P$. 


\section{Analysis}

Figure \ref{fig:polarization}A shows the time-averaged intensity of AIA 193 {\AA} emission, sampling hot 
Fe XXIV emission, from 18:00 - 20:00 UT. During this period, the plasma-sheet dimmed, but with no significant variation to its shape. The plasma-sheet appears as a near-horizontal structure, stretching out from the top of the flare loop arcade (X $\approx 1020$\arcsec). A cross-section through the plasma-sheet places the plasma-sheet centroid at $Y\approx-145$\arcsec\ (Figure \ref{fig:polarization}B).

In comparison, Figures \ref{fig:polarization}C and E respectively show intensity $I$ and linear polarization $P$ of cooler Fe XIII 1074.7 nm emission. All CoMP data shown are ``level 2" data products from an improved pipeline from early October 2019 (de Toma and Galloy, private communication).

The images were calculated using the mean of 46 CoMP $I$, $Q$ and $U$ measurements from 19:44:36 to 20:03:06 UT (later images were excluded due to poorer seeing from passing cloud). Fe XIII 1074.7 nm emission comes from plasma around $\sim$1.5 MK, in contrast to AIA 193 {\AA} at $\sim$1.2 and 20 MK. With clear emission in the post-flare loop-top, the absence of strong Fe XIII intensity in the plasma-sheet is striking. The AIA 211 {\AA} and 193 {\AA} channels do show the plasma-sheet at the 
later times CoMP observed, but emission is weaker and more diffuse than at earlier phases.

The fractional linear polarization $P$ reveals a prominent dark triangular structure with a yet smaller dark structure underneath, just above the limb. The latter feature aligns with the bright flare looptop in the AIA image. The triangular feature however, overlies the bright region over the looptops. These two regions both have $P < 0.01$. Above the dark triangular structure,
aligned roughly along the AIA plasma-sheet emission, there is a broad, dark region, positioned radially from the top of the overlying structure to the west-most edge of the CoMP FOV. A cross section through the region shows a significant drop in polarization (Figure \ref{fig:polarization}F), despite no clear $I$ signature at the same location (Figure \ref{fig:polarization}D).
Here, we see a broad gradual drop in polarization down to $P < 0.01$, from values of $P \approx 0.055$ and $0.075$ either side of the feature. Despite being $\sim 10$ times broader than the structure observed in AIA 193 {\AA}, minimum $P$ occurs at approximately the same location as peak AIA 193 {\AA} emission.

\cite{Cheng} examined the structure in white-light with the K-Cor instrument, measuring the plasma-sheet to be 2.5 times larger in polarized brightness ($pB$) than seen in AIA 193 {\AA}.
This difference may be related to the dependencies of EUV and $pB$ intensities on plasma density $n$ as $n^2$ and $n^1$ respectively. \ion{Fe}{13} emission theoretically depends on $n^{\alpha}$, where $\alpha$ is closer to 1 than 2
owing to radiative excitation
and some collisional depopulation.

Polarization Azimuth angles $\theta$ are shown in Figure \ref{fig:polarization}G. The color map shows angle $\theta$ relative to the local radial, and white lines plot the vectors associated with this angle. Plotted vector length is proportional to $-1/\log(P)$. Polarization vectors are also shown in Figure \ref{fig:polarization}E. The polarization vectors are close to radial above and below the plasma-sheet, and apparently trace the outline of the flare loops and overlaying magnetic field. Beneath this region, azimuth angles are near tangential to the solar
surface. Such behavior is unusual, as azimuth angles normally flip by 90$^\circ$ after crossing $\theta_{VV} \approx 54^\circ$. It can occur under conditions where there is a particular symmetry along the LOS. 

Taking a cross-section of polarization azimuths across the sheet, we see the angles moving from $-20^{\circ}$ to $0^{\circ}$, interrupted by a large dip to $-30^{\circ}$. The peak of the azimuth drop is at the same location as maximum AIA 193 {\AA} emission and minimum \ion{Fe}{13} polarization. This may be a measurement artifact, as noise in $U$ and $Q$ increase as $P$ decreases.

\section{Interpretation}

Linear polarization is created throughout the solar corona. There are only two mechanisms by which linear polarization can be reduced. Firstly, collisions by thermal particles can locally destroy atomic polarization.
Secondly, integrations along the LOS and across the POS can both reduce the net polarization observed, dependent on $\theta_B$. 
At least one of these processes must be responsible for the significant and broad drop in polarization observed across the plasma-sheet.

\subsection{Collisions}

\begin{figure}
  \centering
  \includegraphics[width=8.5cm]{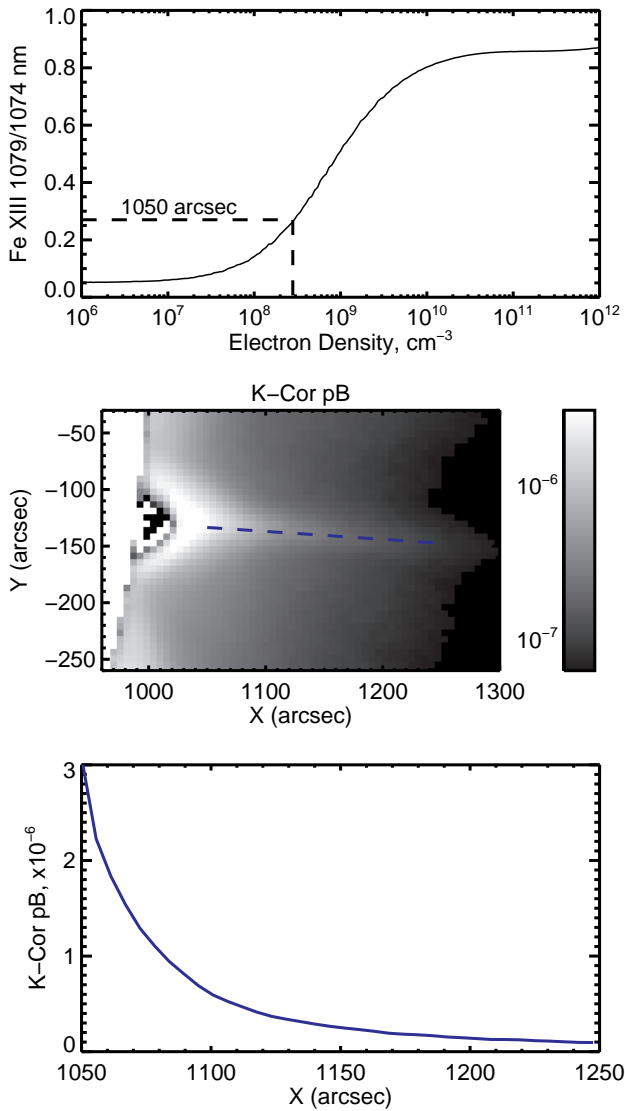}
  \caption{
  Top) Theoretical density curve for Fe XIII 1074.7 to 1079.8 nm intensity ratio. The dashed lines mark the measured ratio at 1050\arcsec\ above the limb, with corresponding density at this location.
  Middle) Time averaged K-Cor polarized brightness ($pB$) observations from 18:00 - 19:30 UT, with units B/Bsun. Blue dashed line marks the location of the cross-section in the panel below.
  Bottom) Variation in $pB$ along the plasma-sheet.
  }
  \label{fig:kcor}
\end{figure}

\begin{figure*}
  \centering
  \includegraphics[width=14.5cm]{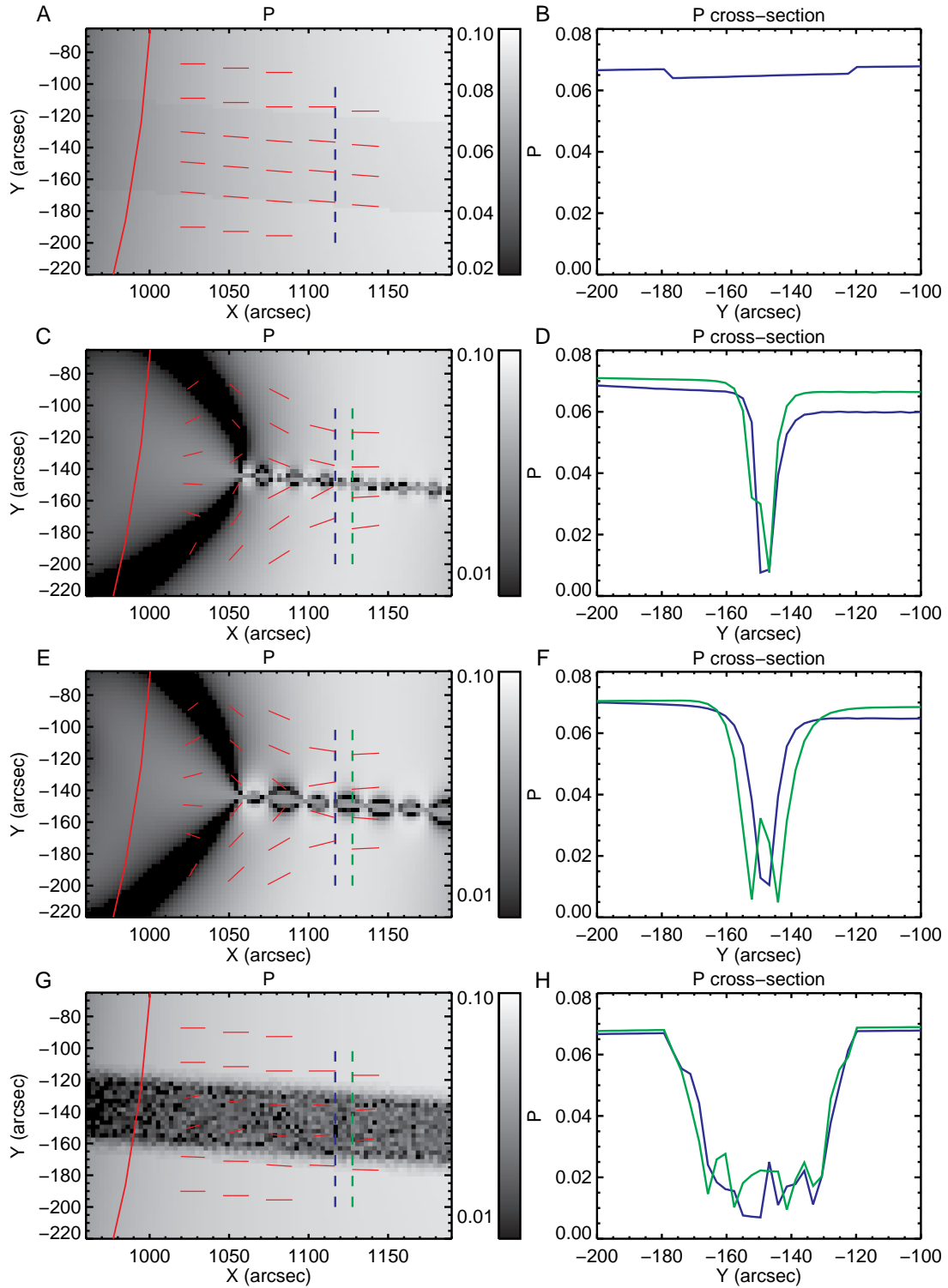}
  \caption{Maps of linear polarization $P$ for plasma-sheet models, with associated cross-sections. CoMP occulting disk position is marked by the red curved line, and polarization vectors as red dashes. Cross-section locations are marked by the blue and green dashed lines, corresponding to the plot color of the cross-section. Models show: 
  AB) Laminar plasma-sheet, with magnetic field parallel to the sheet direction.
  CD) Potential field model generated from infinite LOS line currents placed at \textit{unresolvable} intervals along the plasma-sheet, combined with a sub-surface dipole. This represents plasmoid reconnection within the current-sheet.
  EF) The same potential field model as above, but with currents placed at resolvable distances. 
  GH) Nonphysical plasma-sheet with a randomly orientated field, analogous to plasma turbulence. 
  }
  \label{fig:sim}
\end{figure*}

To determine if collisions are responsible for removing polarization in the plasma-sheet, we must estimate the density of the region. 
The CoMP \ion{Fe}{13} 1074.7 nm and 1079.8 nm lines are a density sensitive pair. We
used the Coronal Line Emission (CLE) program \citep{Judge+Casini2001} to determine the relationship between the line intensity ratio and electron density (at 1.5 MK).
The 1079.8 nm emission is weak, extending only to the base of the plasma-sheet at $\sim1050$\arcsec. At this height, we calculate an electron density at 1.5 MK of $2.8 \times 10^8$ cm$^{-3}$, based on an intensity ratio of 0.27 (Figure \ref{fig:kcor}). 
 
At higher altitudes, this density is likely even lower. We can demonstrate this by measuring the change in total electron density with height, as it is proportional to the polarized brightness $pB$ measured by K-Cor (Figure \ref{fig:kcor}). $n_e$ varies as

\begin{equation}
  n_e \approx \frac{pB} {6.65\times10^{-25} 0.11 \ell},
  \label{eqn:pB}
\end{equation}

\noindent
where $\ell$ is the integration length along the LOS (\citealp{1990ApJ...349..656O}).

At an altitude of 1115\arcsec, $pB=0.5\times10^{-6}$. Therefore, if $\ell$ is greater than the observed plasma-sheet width $w\approx 5$ Mm, $n_e < 1.4\times 10^{10} $ cm$^{-3}$. Assuming $\ell \approx 30$ Mm \citep{Cheng},
$n_e \approx 2\times10^9$ cm$^{-3}$.
These densities are not high enough to destroy atomic polarization via collisions. We therefore conclude that this is unlikely the cause of low linear polarization in the plasma-sheet.

\subsection{Magnetic field structure within the plasma-sheet}
Levels of linear polarization are dependent on magnetic field structure and orientation. The approach of $P$ to zero for angles 
near $\theta_{VV} \approx 54^\circ$ suggests that the small values of $P$ can be accounted for either by a large-scale field close to this angle, or by a more structured field in the POS and/or LOS that contains a mix of angles $\theta_B$.

The morphology of the dark triangular structure in $P$ (Figure \ref{fig:polarization}E) strongly suggests that the Van Vleck effect is operating at the edge of the arcade field, as magnetic field lines wrap around the large-scale current systems producing them. 
Such magnetic null lines are commonly seen in
calculations and data \citep[e.g.][]{Judge+Low+Casini2006,Gibson}.
However, the geometry of a Sweet-Parker plasma-sheet, with a magnetic field direction close to radial, is incompatible with continuous Van Vleck nulls produced in this fashion. Such a configuration is shown in Figure \ref{fig:sim}A, modeled in CLE as an infinitely long laminar current-sheet. Here, we see almost no drop in polarization (Figure \ref{fig:sim}B). Therefore, the levels of $P$ measured by CoMP are inconsistent with a laminar Sweet-Parker current-sheet in the standard eruptive flare model.

Small-scale magnetic structures, such as plasmoids or a turbulent magnetic field (formed perhaps as a result of current-sheet instabilities), naturally lead to variations in $\theta_B$. These structures would therefore cause an overall reduction in $P$, especially if unresolved. To explore such an effect we made simple numerical models using CLE to compare with observations. These simple calculations (Figure \ref{fig:sim}) show polarization levels $P$ and corresponding polarization vectors. Each case utilizes a similar geometry to CoMP observations, with the plasma-sheet centered around the line $y=-0.073$\arcsec$x-67.53$\arcsec. Sample polarization cross-section profiles are also shown. 
The models assume that the plasma contributing mostly to the emission is confined to a narrow region within $10^{-2}R_\odot$ of the POS, to avoid LOS cancellations, and thus to
highlight effects of POS magnetic structure.

Our ``plasmoid'' models (Figures \ref{fig:sim} C\&E) consist of infinite LOS line currents (placed at intervals along the plasma-sheet), combined with the potential field generated from a sub-surface dipole. This 2D configuration is the simplest representation of what might constitute a series of magnetic islands formed by the plasmoid instability in the POS.

Figures \ref{fig:sim}C\&E show calculations of plasmoids with sizes below and above the resolvable limit respectively. In both cases, the interaction between the plasma-sheet edge and surface dipole form a black "V-shaped" structure, where field lines trace an angle close to the Van Vleck angle. The V-shapes
occur here as 
magnetic field lines wrap
around the line currents 
placed along the LOS within the plasma sheet. This is unlikely the cause of the similar structure in CoMP observations however, as the model relies on the infinite line currents to form this feature. Our primary region of interest is the plasma-sheet above this region. 

In the unresolvable plasmoid model (Figure \ref{fig:polarization}C), polarization drops to a minimum of $P\approx0.01$ (Figure \ref{fig:polarization}D). The resolved plasmoid case (Figure \ref{fig:polarization}E) has much more variation in $P$ along the plasma-sheet however, varying greatly between the plasmoid edge and center. In this resolvable case, minimum polarization is calculated as $P\approx0.005$.

In addition to the plasmoid models, we calculated an un-physical model of a randomly orientated field configuration running along the plasma-sheet, again with plasma within $10^{-2}R_\odot$ of the POS 
(Figure \ref{fig:sim}G). Across this structure, we calculate a drop in polarization of to $P\approx 0.015$.
In this case, using a random
field structure imitates the signal of physical fields which, when integrated over finite volumes, contain the same distribution of vector 
magnetic fields. 

Although these models are relatively simple, they provide an analogue for the polarization levels CoMP might observe for representative magnetic topologies. In both the plasmoid and random field cases, magnetic field orientation is shown to be capable of reducing polarization to that observed in this event. With future observations (given an adequate signal-to-noise and integration time), spectropolarimetric measurements can provide observational constraints for the theoretical nature and scale of magnetic substructure in the corona.

\section{Discussion}

In summary, 
polarization data from CoMP seem to demonstrate three properties:
\begin{enumerate}
  \item A broad and gradual reduction of linear polarization 
  across the plasma-sheet, with lowest amplitudes at the sheet center.
  \item No clear increase in IR intensity, in contrast with EUV emission.
  \item Coherent and large `V-shaped' structures of low polarization below the plasma-sheet, reminiscent of the Van Vleck nulls clear in earlier calculations. \citep{Judge+Casini2001}
  \item Near-tangential polarization vectors beneath the plasma-sheet, roughly aligned with the aforementioned dark `V-shaped' structures.
\end{enumerate}

The near-tangential polarization seen under the plasma-sheet is certainly unusual, as near-radial polarization is found far more frequently \citep[e.g.][]{Arnaud}. This is potentially a LOS integration effect through the plasma, canceling out only radial components of polarization. This could perhaps provide information on the large-scale field structure under the plasma-sheet, but is an area of future study and does not effect the conclusions drawn in this paper. 

Although our simple plasmoid models can produce the minimum polarization levels observed by CoMP in the September 10th 2017 flare, they do not replicate the gradual and wide drop in the polarization structure, significantly broader than EUV observations of the plasma-sheet. Recent sophisticated calculations suggest a natural explanation, consistent with the observed behavior of $P$ across the sheet \citep{Stanier}. Cascades of plasmoids caused by fragmentation of finer and finer current-sheets 
diffuse from modeled plasma-sheets much faster than the plasma itself. We might expect this to produce a similar polarization signature to that observed by CoMP in the later phases of
this event. It may also explain in part why the sheet was essentially invisible in 
the measured Fe~XIII intensity, but visible in $P$. 
We speculate that if CoMP had started observing at the start of the flare, we would have observed a polarization structure of similar width to observed intensity, broadening as the process calculated by 
\cite{Stanier} evolves.

\section{Conclusions}

We find that the drop of linear polarization measured by CoMP in the September 10th 2017 flare is consistent with the presence of plasmoids and turbulent fluctuations in the magnetic field. While previous work has focused upon spatially resolvable features in images of the dynamic corona, we have shown that linear polarization
can serve as measure for random magnetic structure on sub-resolution scales. Our method provides a diagnostic to analyze the fragmentation of current sheets through the plasmoid instability, which creates a cascade of energy associated with magnetic-field fluctuations toward smaller scales, well below observable limits.

Our work is consistent with theoretical work suggesting the links between current-sheet dynamics, plasmoid fragmentation, and a turbulent cascade of energy associated with magnetic-field fluctuations. 

We anticipate that the start of observations with DKIST will provide data of the necessary quality to further disentangle the intriguing physics discussed here. In particular, a spectrograph (in contrast to the CoMP filtergraph), could further explore the mystery of why $P$ is clearly
related to the plasma-sheet 
but $I$ is not.

\acknowledgments
R.J.F. thanks the STFC for support via funding from the PhD Studentship, as well as NCAR for funding visits to the High Altitude Observatory via the Newkirk Fellowship. The National Center for Atmospheric Research is sponsored by the National Science Foundation. CoMP data services were provided by MLSO. LvDG and SAM are partially funded under STFC consolidated grant number ST/S000240/1. The authors would also like to thank Daniel Verscharen for his helpful thoughts on theoretical plasma physics.

\bibliographystyle{aasjournal}
\bibliography{bibliography}
\end{document}